\documentclass[10pt, twocolumn, a4paper]{article}

\usepackage[margin=1.7cm, top=2.0cm, bottom=2.0cm]{geometry}
\usepackage{amsmath}
\usepackage{amssymb}
\usepackage{graphicx}
\usepackage{booktabs}
\usepackage{array}
\usepackage{tabularx}
\usepackage{float}
\usepackage{enumitem}
\usepackage{xcolor}
\usepackage{listings}
\usepackage{mdframed}
\usepackage{tikz}
\usepackage{titling}
\usepackage{fancyhdr}
\usepackage{caption}
\usepackage{url}
\usepackage{hyperref}
\usetikzlibrary{arrows.meta, positioning, shapes.geometric, fit, backgrounds, calc}

\hypersetup{
  colorlinks=true, linkcolor=blue!55!black, citecolor=green!50!black,
  urlcolor=blue!55!black,
  pdftitle={A Gateway Architecture for Enterprise MCP Authentication},
  pdfauthor={Suraj Kumar, Amy Wang, Srinivasan Manoharan},
  unicode=true,
}

\providecommand{\keywords}[1]{\smallskip\noindent\textbf{Keywords:} #1}

\setlist{nosep, leftmargin=1.2em}

\lstdefinestyle{httpStyle}{
  basicstyle=\ttfamily\scriptsize, breaklines=true,
  frame=single, rulecolor=\color{gray!40}, backgroundcolor=\color{blue!3},
  numbers=none, tabsize=2, showstringspaces=false, aboveskip=4pt, belowskip=4pt,
}
\lstdefinestyle{jsonStyle}{
  basicstyle=\ttfamily\scriptsize, breaklines=true,
  frame=single, rulecolor=\color{gray!40}, backgroundcolor=\color{gray!5},
  numbers=none, tabsize=2, showstringspaces=false, aboveskip=4pt, belowskip=4pt,
}

\newmdenv[linecolor=blue!30,backgroundcolor=blue!5,roundcorner=3pt,
  innerleftmargin=5pt,innerrightmargin=5pt,innertopmargin=3pt,
  innerbottommargin=3pt,skipabove=4pt,skipbelow=4pt]{infobox}
\newmdenv[linecolor=orange!50,backgroundcolor=orange!6,roundcorner=3pt,
  innerleftmargin=5pt,innerrightmargin=5pt,innertopmargin=3pt,
  innerbottommargin=3pt,skipabove=4pt,skipbelow=4pt]{keybox}

\colorlet{cblue}{blue!70!black}\colorlet{clblue}{blue!8}
\colorlet{cgreen}{green!50!black}\colorlet{clgreen}{green!8}
\colorlet{corange}{orange!75!black}\colorlet{clorange}{orange!10}
\colorlet{cping}{red!60!black}\colorlet{clping}{red!6}
\colorlet{cpurp}{violet!65!black}\colorlet{clpurp}{violet!8}
\colorlet{cgray}{gray!55}

\begin{document}

\title{\bfseries A Gateway Architecture for Enterprise MCP Authentication:\\[3pt]
\large Unifying Heterogeneous Auth, Identity Delegation, and the User /
Non-User Persona Problem}

\author{
   Suraj Kumar\thanks{\texttt{suraj.kumar.bh.in@gmail.com}} \quad
   Amy Wang\thanks{\texttt{yuwewang@paypal.com}} \quad
  Srinivasan Manoharan\thanks{\texttt{srinivmanoharan@paypal.com}} \\[3pt]
  \small Enterprise AI Platform Engineering
}
\date{August 2026}

\maketitle
\thispagestyle{fancy}

\begin{abstract}
The Model Context Protocol (MCP) has become the de-facto interface for connecting LLM
agents to enterprise tools, and adoption has been explosive: within a year, large
organizations went from zero to dozens of internally built MCP servers. That speed
created a governance crisis. Each team implemented authentication independently---some
with no auth, some with API keys, some with full OAuth---producing a fragmented landscape
with no consistent way to authorize callers, track who did what, or offboard a departing
employee across the fleet. This paper reports an industry deployment that resolves the
crisis with a centralized \emph{MCP gateway}: a single aggregation, governance, and
authentication layer that fronts every downstream MCP server.

We make four contributions grounded in production experience. First, we present a
\textbf{two-axis authentication model} that every MCP server must satisfy---a
\emph{persona axis} (interactive \emph{user} vs.\ automated \emph{non-user}) crossed with
a \emph{credential axis} (no-auth, static/dynamic API key, authorization-code-with-PKCE
over static/dynamic clients, client credentials, and platform app-context)---all served
through a single MCP endpoint. Second, we describe the \textbf{gateway authentication
layer} itself, which supports three enterprise SSO grants (Authorization~Code~+~PKCE,
Device~Code, and Resource~Owner~Password~Credentials) and offers callers a choice of
token-provisioning models: Bring-Your-Own-Token (BYOT), Generate-Your-Own-Token (GYOT),
and full delegated OAuth via RFC~8693 token exchange. Third, we generalize
\textbf{three end-to-end identity flows}---User$\rightarrow$OAuth2,
Non-user$\rightarrow$Service-Account, and User$\rightarrow$Service-Account---that compose
client, gateway, and server. Fourth, we document the \textbf{deployment evolution} from a
CDN/WAF/edge perimeter to private MCP tunnels, and the design of enterprise-wide
connectors that let an entire workforce share governed access to AI tools. The
architecture is in production, fronting dozens of MCP servers across web, desktop,
custom-SDK, and low-code clients.
\end{abstract}


\section{Introduction}

\subsection{The Governance Crisis Behind MCP Adoption}

When the Model Context Protocol~\cite{mcp2024} was published in late 2024, it solved a
real problem elegantly: it gave LLM agents a uniform way to discover and invoke external
tools. Adoption inside large enterprises was immediate and uncoordinated. A team that
wanted to expose its service to an internal AI assistant could stand up an MCP server in
an afternoon. Within months, a single organization could find itself running dozens of
internally built MCP servers---for data warehouses, observability platforms, productivity
suites, ticketing systems, and low-code automation engines.

The speed of adoption is exactly what created the problem. The MCP specification says
implementations \emph{should} support OAuth~2.0 but leaves every detail to the
implementer. So every team made its own choice. Some observability servers shipped with
no authentication at all, reasoning that they were ``internal only.'' Others used a single
static API key shared across all callers. A few implemented full OAuth flows. The result
was a fragmented authentication landscape with three concrete failure modes:

\begin{itemize}
  \item \textbf{No consistent authorization.} Each server decided independently who could
    call it and with what privileges. There was no central policy.
  \item \textbf{No unified audit.} Each server logged differently, if at all. Answering
    ``which user invoked which tool last Tuesday, and on whose behalf'' required
    correlating incompatible logs across a dozen systems.
  \item \textbf{No clean offboarding.} When an employee left, revoking their access meant
    contacting every server team individually. There was no single switch.
\end{itemize}

\subsection{A Motivating Example: the Office 365 MCP Server}

\noindent Consider a concrete server that any enterprise will recognize: an Office~365
MCP server exposing mail, calendar, and document tools. It must serve two very different
kinds of caller through \emph{the same endpoint}:

\begin{enumerate}
  \item An \textbf{interactive employee} asking an AI assistant to ``summarize my unread
    mail.'' This call must run \emph{as that user}: it must see only their mailbox, and
    the downstream Microsoft Graph token must carry their identity. This requires an
    interactive OAuth authorization-code flow with the downstream identity provider.
  \item A \textbf{nightly automation} that compiles a team calendar digest. No human is
    in the loop. It authenticates with its own service-account credentials
    (client~credentials), and must \emph{not} be able to impersonate any individual user.
\end{enumerate}

\noindent A naive design would stand up two servers---one per persona---doubling the
operational surface and splitting the tool catalog. Worse, if the automation path is not
explicitly constrained, a misconfiguration lets a headless agent obtain a user-scoped
token, a textbook confused-deputy escalation~\cite{hardy1988}. The Office~365 case is not
special; \emph{every} useful enterprise MCP server faces the same two-persona requirement.
This single example motivates the entire architecture: one endpoint, two personas, many
underlying credential mechanisms, governed centrally.

\subsection{Contributions}

This is an experience paper. The architecture it describes is deployed in production,
fronting dozens of MCP servers used by an entire workforce through shared connectors. Our
contributions are:

\begin{itemize}
  \item \textbf{C1 --- A two-axis authentication model for MCP servers}
    (Section~\ref{sec:twoaxis}). We show that the authentication needs of every enterprise
    MCP server factor cleanly into a \emph{persona} axis (user vs.\ non-user) crossed with
    a \emph{credential} axis (no-auth, static/dynamic API key,
    authorization-code-with-PKCE over static/dynamic clients, client credentials, platform
    app-context), all served through one endpoint.

  \item \textbf{C2 --- A gateway authentication layer with three SSO grants and three
    token-provisioning models} (Section~\ref{sec:gateway}). The gateway authenticates
    callers via Authorization~Code~+~PKCE, Device~Code, or ROPC, and lets integrators
    choose Bring-Your-Own-Token (BYOT), Generate-Your-Own-Token (GYOT), or full delegated
    OAuth via RFC~8693~\cite{rfc8693}. The gateway authenticates; downstream servers verify
    the issued tokens using a shared SDK.

  \item \textbf{C3 --- Three generalized end-to-end identity flows}
    (Section~\ref{sec:flows}): User$\rightarrow$OAuth2, Non-user$\rightarrow$Service-Account,
    and the subtle User$\rightarrow$Service-Account case, each composing client, gateway,
    and server with explicit token handling.

  \item \textbf{C4 --- A deployment and connector evolution report}
    (Section~\ref{sec:deploy}): from a CDN/WAF/edge perimeter with per-vendor IP
    allowlisting to private MCP tunnels, plus the design of enterprise-wide web and desktop
    connectors shared across the workforce.
\end{itemize}

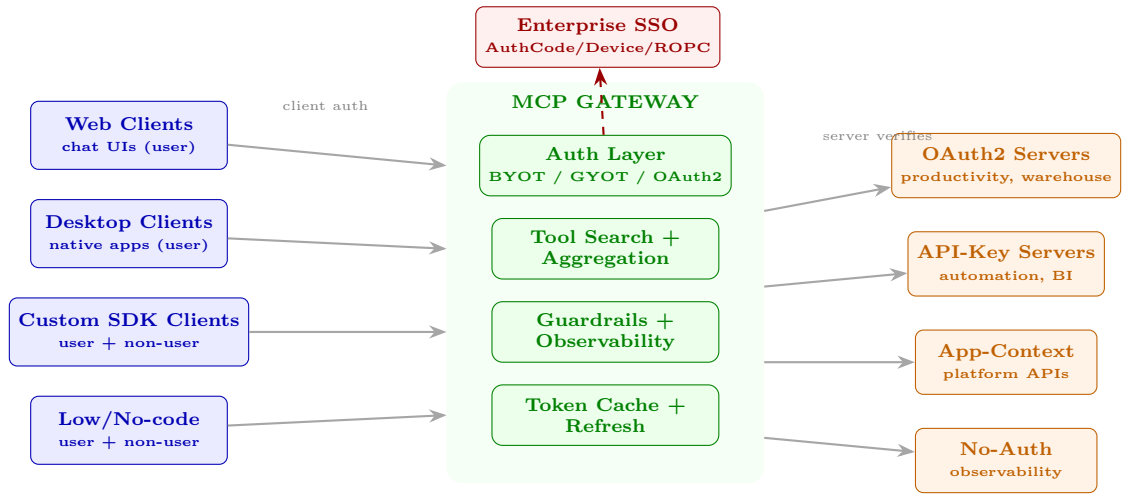
\begin{figure*}[t]
\centering
\begin{tikzpicture}[x=1cm, y=1cm,
  cli/.style={rectangle,rounded corners=3pt,minimum width=2.6cm,minimum height=0.9cm,
    font=\scriptsize\bfseries,align=center,draw=cblue,fill=clblue,text=cblue},
  gw/.style={rectangle,rounded corners=4pt,minimum width=3.0cm,minimum height=0.8cm,
    font=\scriptsize\bfseries,align=center,draw=cgreen,fill=clgreen,text=cgreen},
  srv/.style={rectangle,rounded corners=3pt,minimum width=2.4cm,minimum height=0.85cm,
    font=\scriptsize\bfseries,align=center,draw=corange,fill=clorange,text=corange},
  idp/.style={rectangle,rounded corners=3pt,minimum width=2.4cm,minimum height=0.8cm,
    font=\scriptsize\bfseries,align=center,draw=cping,fill=clping,text=cping},
  arr/.style={-Stealth,thick,gray!70},
]
\node[cli] (c1) at (0, 3.0) {Web Clients\\\tiny chat UIs (user)};
\node[cli] (c2) at (0, 1.7) {Desktop Clients\\\tiny native apps (user)};
\node[cli] (c3) at (0, 0.4) {Custom SDK Clients\\\tiny user + non-user};
\node[cli] (c4) at (0,-0.9) {Low/No-code\\\tiny user + non-user};

\node[idp] (sso) at (6.2, 4.3) {Enterprise SSO\\\tiny AuthCode/Device/ROPC};

\begin{scope}[on background layer]
  \fill[green!4,rounded corners=6pt] (4.2,-1.6) rectangle (8.4,3.7);
  \node[font=\scriptsize\bfseries,text=cgreen] at (6.3,3.45) {MCP GATEWAY};
\end{scope}
\node[gw] (g1) at (6.3, 2.6) {Auth Layer\\\tiny BYOT / GYOT / OAuth2};
\node[gw] (g2) at (6.3, 1.5) {Tool Search +\\Aggregation};
\node[gw] (g3) at (6.3, 0.4) {Guardrails +\\Observability};
\node[gw] (g4) at (6.3,-0.7) {Token Cache +\\Refresh};

\node[srv] (s1) at (11.6, 2.6) {OAuth2 Servers\\\tiny productivity, warehouse};
\node[srv] (s2) at (11.6, 1.3) {API-Key Servers\\\tiny automation, BI};
\node[srv] (s3) at (11.6, 0.0) {App-Context\\\tiny platform APIs};
\node[srv] (s4) at (11.6,-1.3) {No-Auth\\\tiny observability};

\draw[arr] (c1) -- (4.2,2.6);
\draw[arr] (c2) -- (4.2,1.5);
\draw[arr] (c3) -- (4.2,0.4);
\draw[arr] (c4) -- (4.2,-0.7);
\draw[arr,dashed,cping] (g1) -- (sso);
\draw[arr] (8.4,2.0) -- (s1);
\draw[arr] (8.4,1.0) -- (s2);
\draw[arr] (8.4,0.0) -- (s3);
\draw[arr] (8.4,-1.0) -- (s4);

\node[font=\tiny,text=gray] at (2.6,3.4) {client auth};
\node[font=\tiny,text=gray] at (9.9,3.0) {server verifies};
\end{tikzpicture}
\caption{System overview. Heterogeneous clients authenticate \emph{once} to the gateway,
which authenticates them against enterprise SSO and resolves the correct downstream
credential per MCP server. The gateway is the single point for auth, aggregation,
guardrails, observability, and token caching. Downstream servers only \emph{verify}
gateway-issued tokens via a shared SDK.}
\label{fig:bigpicture}
\end{figure*}

\noindent Throughout, we use diagrams as the primary explanatory device: the model is
fundamentally about \emph{which token flows where}, and that is far clearer drawn than
described. Figure~\ref{fig:bigpicture} previews the whole system.

\section{Background and Problem Framing}

\subsection{MCP in Brief}

MCP defines a client--server protocol in which a server exposes \emph{tools} (callable
functions), \emph{resources} (readable data), and \emph{prompts} (templates). The
transport is JSON-RPC~2.0 carried over Streamable~HTTP (the current recommended
transport), HTTP~SSE (the original streaming transport), or stdio for local processes.
Crucially, the protocol specifies \emph{what} a tool call looks like but says nothing
binding about \emph{who} may make it---authentication and authorization are left entirely
to the implementer. That single omission is the root of the governance crisis in
Section~1.1.

\subsection{Why MCP Auth Is Hard: Heterogeneity}

The difficulty is not that authentication is unsolved---it is that enterprise MCP servers
need \emph{many different kinds} of authentication simultaneously, and the right kind
depends on both the downstream system and the calling persona. A data-warehouse server
fronts a system that speaks OAuth; an automation engine fronts one that only accepts a
static API key; an observability server may need no auth at all. The same server must
serve an interactive user (whose downstream calls should run as them) and a headless
automation (which must run as a service account). No single credential type covers this
space. The contribution of Section~\ref{sec:twoaxis} is to show that the space is not
chaotic---it factors into two clean axes.

\subsection{Why a Gateway}

Faced with heterogeneity, an enterprise has two choices: make every server team solve auth
themselves (the status quo that caused the crisis), or centralize. We centralize, using a
\emph{gateway}~\cite{servicemesh} that fronts every MCP server. The gateway aggregates
many servers behind one endpoint (with tool search to avoid context-window bloat from
hundreds of tool schemas), enforces guardrails, emits unified observability and metrics,
caches and refreshes tokens, and---the focus of this paper---provides the authentication
layer that individual servers no longer have to build. Servers are reduced to
\emph{verifying} tokens the gateway issues, using a shared SDK. We compare this against the
alternatives in Table~\ref{tab:perserver}.

\begin{table}[h]
\caption{Per-server auth vs.\ centralized gateway.}
\label{tab:perserver}
\small
\begin{tabularx}{\columnwidth}{p{2.6cm}XX}
\toprule
\textbf{Dimension} & \textbf{Per-server} & \textbf{Gateway} \\
\midrule
SSO registrations  & $N$ (one each) & 1 \\
Audit trail        & Fragmented     & Unified \\
Offboarding        & Contact each team & Central revoke \\
Token caching      & Re-implemented & Shared \\
Tool-context bloat & $N$ catalogs   & Tool search \\
Guardrails         & Per-team       & Centralized \\
\bottomrule
\end{tabularx}
\end{table}

\section{The Two-Axis Authentication Model}
\label{sec:twoaxis}

Our central design claim is that the seemingly chaotic space of enterprise MCP
authentication factors into two orthogonal axes, and that a single MCP endpoint can serve
the entire matrix. The \textbf{persona axis} distinguishes an interactive \emph{user} from
an automated \emph{non-user} (batch jobs, scheduled pipelines, agents with no human in the
loop). The \textbf{credential axis} enumerates the concrete mechanisms. Figure~\ref{fig:twoaxis} shows the full model; the persona determines which credential mechanisms are
\emph{available}, and the downstream system determines which is \emph{used}.

\begin{figure*}[t]
\centering
\begin{tikzpicture}[x=1cm, y=1cm,
  hdr/.style={rectangle,rounded corners=3pt,minimum height=0.6cm,
    font=\scriptsize\bfseries,align=center},
  leaf/.style={rectangle,rounded corners=3pt,minimum width=3.7cm,minimum height=0.62cm,
    font=\scriptsize,align=center,draw,fill=white},
  cat/.style={rectangle,rounded corners=3pt,minimum width=1.9cm,minimum height=0.58cm,
    font=\scriptsize\bfseries,align=center},
]
\node[hdr,fill=clblue,draw=cblue,text=cblue,minimum width=7.0cm] (uh) at (3.3, 6.2)
  {USER PERSONA \ \tiny(interactive, human in the loop)};

\node[cat,fill=clblue!60,draw=cblue,text=cblue] (un)  at (0.2, 5.1) {No Auth};
\node[cat,fill=clblue!60,draw=cblue,text=cblue] (uk)  at (0.2, 4.0) {API Key};
\node[cat,fill=clblue!60,draw=cblue,text=cblue] (uc)  at (0.2, 2.3) {Auth Code\\+ PKCE};

\node[leaf] (un1) at (4.3, 5.1) {observability platform};
\node[leaf] (uk1) at (4.3, 4.45){Static key \ \tiny(automation engine)};
\node[leaf] (uk2) at (4.3, 3.55){Dynamic key \ \tiny(BI / analytics)};
\node[leaf] (uc1) at (4.3, 2.75){Static client \ \tiny(productivity, warehouse)};
\node[leaf] (uc2) at (4.3, 1.85){Dynamic client \ \tiny(ticketing, observability)};

\draw[-Stealth,cblue] (un) -- (un1);
\draw[-Stealth,cblue] (uk) -- (uk1);
\draw[-Stealth,cblue] (uk) -- (uk2);
\draw[-Stealth,cblue] (uc) -- (uc1);
\draw[-Stealth,cblue] (uc) -- (uc2);

\draw[gray!45,dashed,thick] (7.4,6.5) -- (7.4,1.2);

\node[hdr,fill=clorange,draw=corange,text=corange,minimum width=7.0cm] (nh) at (11.5, 6.2)
  {NON-USER PERSONA \ \tiny(batch / automation, no human)};

\node[cat,fill=clorange!60,draw=corange,text=corange] (nn) at (8.6, 5.1) {No Auth};
\node[cat,fill=clorange!60,draw=corange,text=corange] (nk) at (8.6, 4.0) {API Key};
\node[cat,fill=clorange!60,draw=corange,text=corange] (ncc)at (8.6, 2.6) {Client\\Creds};
\node[cat,fill=clorange!60,draw=corange,text=corange] (nac)at (8.6, 1.5) {App\\Context};

\node[leaf] (nn1) at (12.7, 5.1) {observability platform};
\node[leaf] (nk1) at (12.7, 4.45){Static key \ \tiny(automation, ticketing)};
\node[leaf] (nk2) at (12.7, 3.55){Dynamic key \ \tiny(BI / analytics)};
\node[leaf] (nc1) at (12.7, 2.6) {Client creds \ \tiny(productivity suite)};
\node[leaf] (na1) at (12.7, 1.5) {App-context \ \tiny(ops tooling)};

\draw[-Stealth,corange] (nn) -- (nn1);
\draw[-Stealth,corange] (nk) -- (nk1);
\draw[-Stealth,corange] (nk) -- (nk2);
\draw[-Stealth,corange] (ncc)-- (nc1);
\draw[-Stealth,corange] (nac)-- (na1);

\node[font=\scriptsize\itshape,text=gray] at (7.4,0.5)
  {Both personas reach the SAME MCP server through ONE endpoint};
\end{tikzpicture}
\caption{The two-axis authentication model. Each MCP server supports both personas through
a single endpoint. The persona (user vs.\ non-user) gates which credential mechanisms are
available; the downstream system determines which is actually used. Examples in
parentheses are representative server classes.}
\label{fig:twoaxis}
\end{figure*}

\subsection{User Persona}

An interactive user is a human driving an AI client. The server can authenticate them in
one of three ways, in increasing order of identity fidelity:

\begin{itemize}
  \item \textbf{No auth:} used only for low-sensitivity observability platforms where the
    data carries no per-user authorization. The gateway still records the caller; the
    \emph{server} simply does not require a credential.
  \item \textbf{API key:} either a \textbf{static} key (a single long-lived secret, e.g.\
    a low-code automation engine) or a \textbf{dynamic} key constructed per-user from
    stored credentials (e.g.\ a BI/analytics platform that mints a short-lived key bound
    to the requesting user).
  \item \textbf{Authorization code with PKCE:} the high-fidelity path---an interactive
    OAuth flow that produces a downstream token carrying the user's identity. The
    downstream OAuth client may be \textbf{static} (one shared client registration, e.g.\
    a productivity suite or data warehouse) or \textbf{dynamic} (a client registered
    per tenant or per integration, typical of multi-tenant SaaS such as a ticketing
    platform or an observability platform that issues each customer its own OAuth client).
\end{itemize}

\subsection{Non-User Persona}

A non-user caller is an automation with no human present. Its available mechanisms differ
because there is no interactive consent step and no user identity to carry:

\begin{itemize}
  \item \textbf{No auth:} the same low-sensitivity observability case.
  \item \textbf{API key:} \textbf{static} (automation, ticketing, or observability
    platforms that accept a service key) or \textbf{dynamic} (a BI/analytics platform that
    mints a short-lived key for the service account).
  \item \textbf{Client credentials:} the OAuth machine-to-machine grant, used where the
    downstream system speaks OAuth and can accept a service-account token (e.g.\ a
    productivity suite via its directory's client-credentials grant).
  \item \textbf{Platform app-context:} a platform-native machine identity (e.g.\ an
    internal app-context credential for ops/SRE tooling) where the downstream system trusts
    a derived application context rather than a bearer token.
\end{itemize}

\begin{keybox}
\textbf{Key insight (C1).} The persona axis is not cosmetic---it is a \emph{security
boundary}. Authorization-code-with-PKCE is available to users (it requires interactive
consent) but \emph{not} to non-users; client credentials and app-context are available to
non-users but never carry a user identity. This asymmetry is what prevents a headless
automation from impersonating a human.
\end{keybox}

\subsection{One Endpoint, Both Personas}

The architectural commitment is that a single MCP server endpoint serves both personas.
The server does not run two ports or two catalogs. Instead, the \emph{token presented}
determines the persona, and the gateway---not the server---resolves the appropriate
downstream credential. Table~\ref{tab:matrix} summarizes which mechanisms each persona may
use.

\begin{table}[h]
\caption{Credential mechanisms available per persona. \checkmark{}~= available;
\textemdash{}~= not available by design.}
\label{tab:matrix}
\small
\begin{tabularx}{\columnwidth}{Xcc}
\toprule
\textbf{Credential mechanism} & \textbf{User} & \textbf{Non-user} \\
\midrule
No auth                       & \checkmark & \checkmark \\
Static API key                & \checkmark & \checkmark \\
Dynamic API key               & \checkmark & \checkmark \\
Auth code + PKCE (static client)  & \checkmark & \textemdash \\
Auth code + PKCE (dynamic client) & \checkmark & \textemdash \\
Client credentials            & \textemdash & \checkmark \\
Platform app-context          & \textemdash & \checkmark \\
\bottomrule
\end{tabularx}
\end{table}

\section{The Gateway Authentication Layer}
\label{sec:gateway}

\subsection{Design Evolution: How We Got Here}

The gateway's auth design was not chosen up front; it evolved as the deployment grew.
Telling that story is the clearest way to motivate the final design.

\textbf{Stage 0 --- the user flow only.} The first deployments served interactive users
exclusively. People wanted an AI assistant that could read their data \emph{as them}, so
the obvious first build was an interactive OAuth flow: the user logs in through the
enterprise SSO, and the gateway obtains a token on their behalf. This worked, and for a
while the only persona was the human user.

\textbf{Stage 1 --- the need for automation.} Demand quickly arrived for non-interactive
usage: nightly digests, batch enrichment, agentic workflows that run unattended. These
have no browser and no human to complete a consent screen. The interactive flow could not
serve them. We added a machine path---service-account credentials---and with it the
\emph{non-user} persona was born. This is the moment the two-axis model became necessary.

\textbf{Stage 2 --- heterogeneity bites.} As more servers came online, we hit the
heterogeneity wall of Section~2.2: different downstream systems demanded different
credential types. Rather than let each server reinvent auth, we pushed all of it into the
gateway and reduced servers to token \emph{verifiers}.

\begin{figure}[h]
\centering
\begin{tikzpicture}[x=1cm,y=1cm,
  st/.style={rectangle,rounded corners=3pt,minimum width=1.7cm,minimum height=1.0cm,
    font=\tiny\bfseries,align=center,draw,fill=white},
  arr/.style={-Stealth,thick,gray!60},
]
\node[st,draw=cblue,fill=clblue,text=cblue]   (s0) at (0,0)   {Stage 0\\User flow\\(OAuth)};
\node[st,draw=corange,fill=clorange,text=corange](s1) at (2.1,0) {Stage 1\\Non-user\\(SA creds)};
\node[st,draw=cgreen,fill=clgreen,text=cgreen] (s2) at (4.2,0) {Stage 2\\Gateway\\centralize};
\node[st,draw=cpurp,fill=clpurp,text=cpurp]    (s3) at (6.3,0) {Stage 3\\Delegation\\(RFC 8693)};
\draw[arr] (s0)--(s1); \draw[arr] (s1)--(s2); \draw[arr] (s2)--(s3);
\end{tikzpicture}
\caption{Design evolution. Each stage added a requirement the previous design could not
meet, driving toward the final centralized, delegation-aware gateway.}
\label{fig:evolution}
\end{figure}
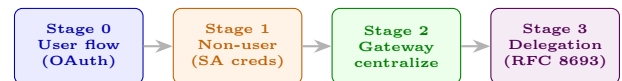

\textbf{Stage 3 --- delegation.} Finally, agentic clients made identity delegation
concrete: a user asks an agent to act, the agent calls the gateway, and the audit log must
show \emph{both} the user and the agent. Standard OAuth carries one subject; we adopted
RFC~8693 token exchange to carry both. Figure~\ref{fig:evolution} summarizes the arc.

\subsection{Three Enterprise SSO Grants}

The gateway authenticates callers against the enterprise SSO using three OAuth grant types,
matched to how the caller runs:

\begin{itemize}
  \item \textbf{Authorization Code + PKCE} --- for \emph{user} personas with a browser
    (web and desktop clients). The standard interactive flow; PKCE~\cite{rfc7636} protects
    the authorization code in transit and is mandatory for all clients.
  \item \textbf{Device Code} --- for \emph{user} personas on input-constrained or remote
    machines (e.g.\ a CLI on a headless dev box). The user authorizes on a second device;
    the originating machine polls for the token.
  \item \textbf{Resource Owner Password Credentials (ROPC)} --- for \emph{non-user}
    personas: service accounts that present credentials directly. ROPC is restricted to
    machine identities and never used for interactive humans.
\end{itemize}

\begin{table}[h]
\caption{SSO grant selection by client and persona.}
\label{tab:grants}
\small
\begin{tabularx}{\columnwidth}{Xll}
\toprule
\textbf{Caller} & \textbf{Persona} & \textbf{Grant} \\
\midrule
Web / desktop client      & User     & Auth Code + PKCE \\
Remote / headless machine & User     & Device Code \\
Service account           & Non-user & ROPC \\
\bottomrule
\end{tabularx}
\end{table}

\subsection{Three Token-Provisioning Models}

A key flexibility of the gateway is that integrators are not forced into one token model.
We support three, covering the full spectrum from ``the client already has a token'' to
``the gateway brokers everything'':

\begin{itemize}
  \item \textbf{BYOT (Bring Your Own Token).} The client already holds a valid SSO token
    and presents it. The gateway validates via introspection and uses it. Lowest friction
    when the client is already SSO-integrated.
  \item \textbf{GYOT (Generate Your Own Token).} The client asks the gateway to initiate a
    grant on its behalf (e.g.\ start a device-code flow), and the gateway returns the
    resulting token. Useful for clients that can drive a flow but do not want to implement
    OAuth themselves.
  \item \textbf{Full delegated OAuth (RFC~8693).} For agent-with-user scenarios, the agent
    exchanges its user-scoped token for a gateway-scoped token in which the SSO injects the
    \texttt{act} (actor) claim. This is the only model that preserves both identities; we
    detail it in Section~\ref{sec:flows}.
\end{itemize}

\begin{keybox}
\textbf{Key insight (C2).} Offering BYOT, GYOT, and full OAuth side by side is what lets a
single gateway serve clients ranging from fully SSO-native web apps to thin custom-SDK
agents---without forcing any of them to re-implement the others' machinery.
\end{keybox}

\subsection{Two Ways to Obtain a Static-Client Token}

The authorization-code-with-PKCE static-client case (e.g.\ a productivity suite, data
warehouse, or SRE server) deserves special note because the gateway supports \emph{two}
distinct ways to obtain the downstream token:

\begin{enumerate}
  \item \textbf{Direct token exchange via the browser path.} The user completes an
    interactive flow against the downstream IdP directly through the browser.
  \item \textbf{Token exchange without the browser path.} The gateway takes the user's
    existing SSO token (already obtained via Authorization~Code~+~PKCE) and performs an
    RFC~8693 token exchange to mint the downstream token---no second browser round-trip.
\end{enumerate}

\begin{figure}[h]
\centering
\begin{tikzpicture}[x=1cm,y=1cm,
  n/.style={rectangle,rounded corners=3pt,minimum width=1.7cm,minimum height=0.7cm,
    font=\tiny\bfseries,align=center,draw,fill=white},
  arr/.style={-Stealth,thick},
]
\node[font=\tiny\bfseries,text=cblue] at (0,2.5) {Path 1: browser};
\node[n,draw=cblue,fill=clblue,text=cblue]   (a1) at (0,1.8) {User\\browser};
\node[n,draw=cping,fill=clping,text=cping]   (a2) at (2.3,1.8) {Downstream\\IdP};
\node[n,draw=corange,fill=clorange,text=corange](a3) at (4.6,1.8) {Downstream\\token};
\draw[arr,cblue] (a1)--(a2); \draw[arr,cping](a2)--(a3);

\node[font=\tiny\bfseries,text=cgreen] at (0,0.6) {Path 2: exchange};
\node[n,draw=cgreen,fill=clgreen,text=cgreen] (b1) at (0,-0.1){SSO token\\(have it)};
\node[n,draw=cping,fill=clping,text=cping]    (b2) at (2.3,-0.1){RFC 8693\\exchange};
\node[n,draw=corange,fill=clorange,text=corange](b3) at (4.6,-0.1){Downstream\\token};
\draw[arr,cgreen](b1)--(b2); \draw[arr,cping](b2)--(b3);
\node[font=\tiny,text=gray] at (2.3,-0.9){no browser round-trip};
\end{tikzpicture}
\caption{Two ways to obtain a static-client downstream token. Path 2 reuses the user's
existing SSO token via RFC~8693, avoiding a redundant browser consent.}
\label{fig:twopath}
\end{figure}
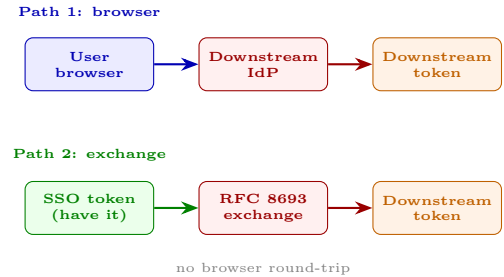

\noindent The second path is strictly better for user experience when the user is already
authenticated to the gateway: it removes a redundant consent screen while preserving full
identity fidelity. Figure~\ref{fig:twopath} contrasts the two.

\subsection{Gateway Issues, Server Verifies}

A clean separation of duties underpins the whole design: the \textbf{gateway
authenticates} and the \textbf{downstream server verifies}. Servers do not implement OAuth;
they include a shared SDK that validates gateway-issued tokens (signature, audience,
expiry, and introspection). The gateway also handles token \textbf{caching} and
\textbf{refresh} so that downstream systems are not hammered with introspection calls and
users are not repeatedly prompted. This is what makes onboarding a new MCP server a matter
of registering it and adding the SDK, rather than building an identity stack.

\section{Three End-to-End Identity Flows}
\label{sec:flows}

The two-axis model (Section~\ref{sec:twoaxis}) and the gateway layer
(Section~\ref{sec:gateway}) compose into three end-to-end flows that span client, gateway,
and downstream server. We present them generalized from a concrete data-warehouse server,
but they apply to any MCP server. The three flows are distinguished by \emph{whose
identity reaches the downstream system}.

\subsection{Flow 1 --- User $\rightarrow$ OAuth2}

A human user runs as themselves end-to-end. The client completes an interactive SSO
authorization-code flow. Crucially, the gateway does \emph{not} hand the raw OAuth2 token
back to the client; instead it stores the token internally and returns an opaque
\emph{session handle}. On every subsequent MCP call the client presents the session, and
the gateway looks up (and silently refreshes, if expired) the token mapped to that session
before validating it and forwarding the call. Keeping the token server-side means it never
transits the client and can be refreshed without re-prompting the user.

\begin{lstlisting}[style=jsonStyle, caption={Flow 1 --- the OAuth2 token the gateway holds (mapped to the session). \texttt{act} is absent because the user authenticated directly.}]
{
  "sub":   "user@corp.com",
  "aud":   "https://gateway.example.com",
  "scope": "mcp:tools",
  "act":   null              // absent: direct user, no delegation
}
\end{lstlisting}

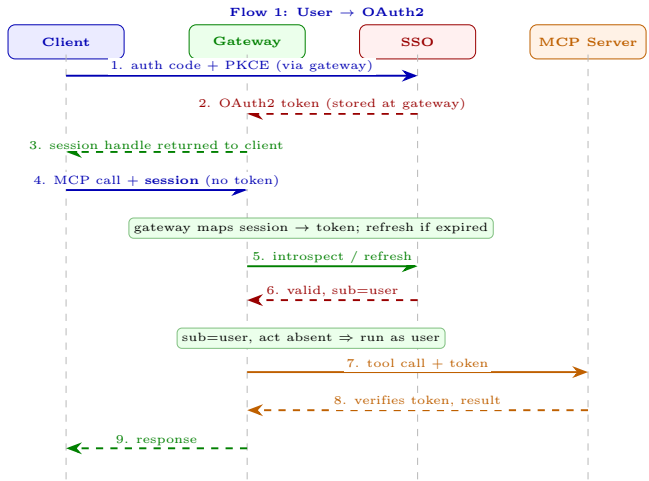
\begin{figure}[h]
\centering
\resizebox{\columnwidth}{!}{%
\begin{tikzpicture}[x=0.78cm,y=0.62cm,
  a/.style={rectangle,rounded corners=3pt,minimum width=1.7cm,minimum height=0.5cm,
    font=\tiny\bfseries,align=center},
  cl/.style={a,fill=clblue,draw=cblue,text=cblue},
  gw/.style={a,fill=clgreen,draw=cgreen,text=cgreen},
  pf/.style={a,fill=clping,draw=cping,text=cping},
  ds/.style={a,fill=clorange,draw=corange,text=corange},
  ll/.style={draw=cgray,dashed,thin}, fw/.style={-Stealth,thick},
  bk/.style={-Stealth,thick,dashed}, lbl/.style={font=\tiny,fill=white,inner sep=1pt,align=center},
  nb/.style={rectangle,rounded corners=2pt,font=\tiny,align=center,inner sep=2pt,draw=cgreen!50,fill=clgreen},
]
\node[cl](C) at (0,-0.3)   {Client};
\node[gw](N) at (3.4,-0.3) {Gateway};
\node[pf](P) at (6.6,-0.3) {SSO};
\node[ds](D) at (9.8,-0.3) {MCP Server};
\foreach \x in {0,3.4,6.6,9.8} \draw[ll](\x,-0.6)--(\x,-10.8);
\draw[fw,cblue](0,-1.1)--(6.6,-1.1) node[lbl,midway,above]{1. auth code + PKCE (via gateway)};
\draw[bk,cping](6.6,-2.0)--(3.4,-2.0) node[lbl,midway,above]{2. OAuth2 token (stored at gateway)};
\draw[bk,cgreen](3.4,-2.9)--(0,-2.9) node[lbl,midway,above]{3. session handle returned to client};
\draw[fw,cblue](0,-3.8)--(3.4,-3.8) node[lbl,midway,above]{4. MCP call + \textbf{session} (no token)};
\node[nb] at (4.6,-4.7){gateway maps session $\rightarrow$ token; refresh if expired};
\draw[fw,cgreen](3.4,-5.6)--(6.6,-5.6) node[lbl,midway,above]{5. introspect / refresh};
\draw[bk,cping](6.6,-6.4)--(3.4,-6.4) node[lbl,midway,above]{6. valid, sub=user};
\node[nb] at (4.6,-7.3){sub=user, act absent $\Rightarrow$ run as user};
\draw[fw,corange](3.4,-8.1)--(9.8,-8.1) node[lbl,midway,above]{7. tool call + token};
\draw[bk,corange](9.8,-9.0)--(3.4,-9.0) node[lbl,midway,above]{8. verifies token, result};
\draw[bk,cgreen](3.4,-9.9)--(0,-9.9) node[lbl,midway,above]{9. response};
\node[font=\tiny\bfseries,text=cblue] at (4.9,0.4){Flow 1: User $\rightarrow$ OAuth2};
\end{tikzpicture}}
\caption{Flow 1. The user authenticates interactively. The gateway keeps the OAuth2 token
server-side and returns an opaque \emph{session} to the client (step~3). Subsequent calls
carry the session, not the token; the gateway resolves and refreshes the mapped token
(step~4--5) before forwarding. The user's identity is preserved end-to-end.}
\label{fig:flow1}
\end{figure}

\subsection{Flow 2 --- Non-user $\rightarrow$ Service Account}

An automation runs as a service account. The agent authenticates via ROPC and presents
\emph{only} its service-account credential on the MCP call. The downstream API key is
\textbf{not} passed by the client---it is \emph{provisioned to the gateway ahead of time}
(out of band, during server onboarding) and the gateway attaches it when forwarding. No
user identity is involved at any point, and the downstream secret never transits the
client.

\begin{figure}[h]
\centering
\resizebox{\columnwidth}{!}{%
\begin{tikzpicture}[x=0.82cm,y=0.62cm,
  a/.style={rectangle,rounded corners=3pt,minimum width=1.7cm,minimum height=0.5cm,
    font=\tiny\bfseries,align=center},
  ag/.style={a,fill=clorange,draw=corange,text=corange},
  gw/.style={a,fill=clgreen,draw=cgreen,text=cgreen},
  pf/.style={a,fill=clping,draw=cping,text=cping},
  ds/.style={a,fill=clorange!60!clblue,draw=orange!50!blue,text=orange!60!blue},
  ll/.style={draw=cgray,dashed,thin}, fw/.style={-Stealth,thick},
  bk/.style={-Stealth,thick,dashed}, lbl/.style={font=\tiny,fill=white,inner sep=1pt,align=center},
  nb/.style={rectangle,rounded corners=2pt,font=\tiny,align=center,inner sep=2pt,draw=cgreen!50,fill=clgreen},
  cfg/.style={rectangle,rounded corners=2pt,font=\tiny,align=center,inner sep=2pt,draw=gray!60,fill=gray!10},
]
\node[ag](A) at (0,-0.3)   {Agent};
\node[gw](N) at (3.6,-0.3) {Gateway};
\node[pf](P) at (7.0,-0.3) {SSO};
\node[ds](D) at (10.4,-0.3){MCP Server};
\foreach \x in {0,3.6,7.0,10.4} \draw[ll](\x,-0.6)--(\x,-8.6);
\node[cfg] at (5.0,-1.0){\textit{Setup (out of band):} downstream API key pre-configured at gateway};
\draw[fw,corange](0,-1.9)--(7.0,-1.9) node[lbl,midway,above]{1. ROPC (service account)};
\draw[bk,cping](7.0,-2.8)--(0,-2.8) node[lbl,midway,above]{2. machine token};
\draw[fw,corange](0,-3.7)--(3.6,-3.7) node[lbl,midway,above]{3. MCP call + machine token \textit{(no key)}};
\node[nb] at (5.0,-4.6){sub=service-account, no act $\Rightarrow$ run as SA;\\attach pre-configured key};
\draw[fw,ds,draw=orange!50!blue](3.6,-5.6)--(10.4,-5.6) node[lbl,midway,above]{4. tool call + pre-configured key};
\draw[bk,draw=orange!50!blue](10.4,-6.5)--(3.6,-6.5) node[lbl,midway,above]{5. verifies key, result};
\draw[bk,cgreen](3.6,-7.4)--(0,-7.4) node[lbl,midway,above]{6. response};
\node[font=\tiny\bfseries,text=corange] at (5.2,0.4){Flow 2: Non-user $\rightarrow$ Service Account};
\end{tikzpicture}}
\caption{Flow 2. A headless agent authenticates via ROPC and sends only its machine token.
The downstream API key is provisioned to the gateway ahead of time (grey note), not passed
on the call; the gateway attaches it when forwarding. No user identity is present and the
downstream secret never reaches the client.}
\label{fig:flow2}
\end{figure}
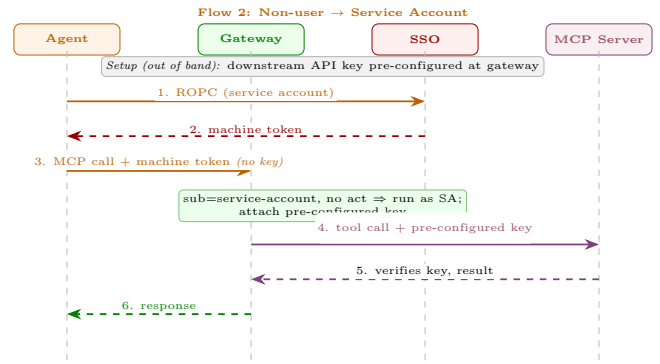

\subsection{Flow 3 --- User $\rightarrow$ Service Account}

The subtle case. A human user is present, but the downstream system is accessed via a
service account---\emph{provided the user is authorized to use that service account}. The
client completes an interactive SSO flow (so the gateway knows the real user), the gateway
\emph{validates that the user has service-account access}, and only then forwards the call
under the service-account credential. This composes user accountability with
service-account execution, and the authorization check is the crux.

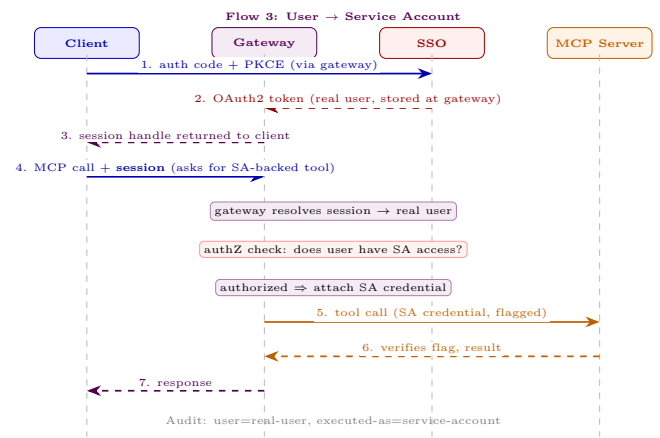
\begin{figure}[h]
\centering
\resizebox{\columnwidth}{!}{%
\begin{tikzpicture}[x=0.80cm,y=0.62cm,
  a/.style={rectangle,rounded corners=3pt,minimum width=1.7cm,minimum height=0.5cm,
    font=\tiny\bfseries,align=center},
  cl/.style={a,fill=clblue,draw=cblue,text=cblue},
  gw/.style={a,fill=clpurp,draw=cpurp,text=cpurp},
  pf/.style={a,fill=clping,draw=cping,text=cping},
  ds/.style={a,fill=clorange,draw=corange,text=corange},
  ll/.style={draw=cgray,dashed,thin}, fw/.style={-Stealth,thick},
  bk/.style={-Stealth,thick,dashed}, lbl/.style={font=\tiny,fill=white,inner sep=1pt,align=center},
  nb/.style={rectangle,rounded corners=2pt,font=\tiny,align=center,inner sep=2pt,draw=cpurp!50,fill=clpurp},
  nbr/.style={rectangle,rounded corners=2pt,font=\tiny,align=center,inner sep=2pt,draw=red!40,fill=red!5},
]
\node[cl](C) at (0,-0.3)   {Client};
\node[gw](N) at (3.6,-0.3) {Gateway};
\node[pf](P) at (7.0,-0.3) {SSO};
\node[ds](D) at (10.4,-0.3){MCP Server};
\foreach \x in {0,3.6,7.0,10.4} \draw[ll](\x,-0.6)--(\x,-10.6);
\draw[fw,cblue](0,-1.1)--(7.0,-1.1) node[lbl,midway,above]{1. auth code + PKCE (via gateway)};
\draw[bk,cping](7.0,-2.0)--(3.6,-2.0) node[lbl,midway,above]{2. OAuth2 token (real user, stored at gateway)};
\draw[bk,cpurp](3.6,-2.9)--(0,-2.9) node[lbl,midway,above]{3. session handle returned to client};
\draw[fw,cblue](0,-3.8)--(3.6,-3.8) node[lbl,midway,above]{4. MCP call + \textbf{session} (asks for SA-backed tool)};
\node[nb] at (5.0,-4.7){gateway resolves session $\rightarrow$ real user};
\node[nbr] at (5.0,-5.7){authZ check: does user have SA access?};
\node[nb]  at (5.0,-6.7){authorized $\Rightarrow$ attach SA credential};
\draw[fw,corange](3.6,-7.6)--(10.4,-7.6) node[lbl,midway,above]{5. tool call (SA credential, flagged)};
\draw[bk,corange](10.4,-8.5)--(3.6,-8.5) node[lbl,midway,above]{6. verifies flag, result};
\draw[bk,cpurp](3.6,-9.4)--(0,-9.4) node[lbl,midway,above]{7. response};
\node[font=\tiny,text=gray] at (5.0,-10.2){Audit: user=real-user, executed-as=service-account};
\node[font=\tiny\bfseries,text=cpurp] at (5.2,0.4){Flow 3: User $\rightarrow$ Service Account};
\end{tikzpicture}}
\caption{Flow 3. A real user authenticates; the gateway stores the token and returns a
session (step~3). The client passes the session (step~4); the gateway resolves the real
user, then performs the user-has-SA-access check \emph{before} attaching the service-account
credential. The audit record captures both the human and the execution identity.}
\label{fig:flow3}
\end{figure}

\begin{keybox}
\textbf{Key insight (C3).} Flow~3 is where most naive designs fail. Letting a user trigger
a service-account-backed tool without an explicit authorization check is the confused-deputy
escalation in disguise. The gateway's user-has-SA-access check, performed \emph{before}
attaching the service-account credential, is the control that makes this flow safe.
\end{keybox}

\subsection{Delegation: Preserving Both Identities}

When the caller is an \emph{agent acting for a user}, Flow~1 is extended with RFC~8693
token exchange. The agent first obtains a user-scoped token (AT1), then exchanges it for a
gateway-scoped token (AT2) in which the SSO injects an \texttt{act} claim naming the agent.
The result carries both identities in one signed token:

\begin{lstlisting}[style=jsonStyle, caption={Delegated token (AT2) --- both identities in one signed JWT. \texttt{sub} is the user (preserved from AT1); \texttt{act.sub} is the agent (injected by the SSO during exchange).}]
{
  "sub":   "user@corp.com",        // user, preserved
  "act": {
    "sub": "agent-prod"            // agent, injected by SSO
  },
  "aud":   "https://gateway.example.com",
  "scope": "mcp:tools"
}
\end{lstlisting}

\noindent The critical configuration detail---and the single most common
misconfiguration---is that the subject-token validation must check the audience of AT1
(the agent's own client ID), \emph{not} the gateway audience. AT1 is scoped to the agent
application; the gateway is the \emph{target} of the exchange, not the current audience.
Multi-agent chains extend this automatically: each hop appends one level of \texttt{act}
nesting, producing a complete, cryptographically bound delegation trail in every token
that reaches the gateway.

\section{Authorization Across Caller Types}
\label{sec:authz}

Authentication establishes \emph{who} is calling; authorization decides \emph{whether they
may}. A recurring source of confusion in MCP deployments is conflating the two. Our design
keeps them distinct and places each authorization decision at a well-defined point. This
section makes the authorization model explicit across the cases of
Sections~\ref{sec:twoaxis}--\ref{sec:flows}.

\subsection{Two Enforcement Points}

Authorization is enforced at two layers, and both matter:

\begin{itemize}
  \item \textbf{Gateway authorization} --- coarse-grained, identity-level decisions made
    before the call is forwarded: is the caller permitted to use this server at all; for a
    user, do they hold the required role; for the User$\rightarrow$SA case, are they
    entitled to the service account.
  \item \textbf{Downstream authorization} --- fine-grained, resource-level decisions made
    by the downstream system using the forwarded credential: which records, which mailbox,
    which dataset. The gateway never tries to replicate this; it forwards an appropriately
    scoped credential and lets the system of record decide.
\end{itemize}

\begin{keybox}
\textbf{Design rule.} The gateway answers ``\emph{may this identity reach this tool?}'';
the downstream system answers ``\emph{may this identity touch this data?}'' Splitting the
decision this way keeps the gateway free of per-resource policy while still enforcing a
hard identity boundary.
\end{keybox}

\subsection{Decision by Caller Type}

Table~\ref{tab:authz} summarizes who decides and what is checked for each caller type. The
asymmetry from Section~\ref{sec:twoaxis} reappears here as an authorization rule, not just
an availability rule: a non-user identity can never satisfy a user-scoped authorization
check, because it carries no user identity to check.

\begin{table}[h]
\caption{Authorization decision points by caller type.}
\label{tab:authz}
\small
\begin{tabularx}{\columnwidth}{p{1.9cm}p{2.2cm}X}
\toprule
\textbf{Caller} & \textbf{Gateway checks} & \textbf{Downstream checks} \\
\midrule
Direct user & user role / scope for the tool & per-resource, as the user \\
\addlinespace
Agent-with-user & user role (agent inherits, never exceeds) & per-resource, as the user \\
\addlinespace
Autonomous agent & service-account role; user-scoped patterns denied & per-resource, as the service account \\
\addlinespace
User $\rightarrow$ SA & \textbf{user-has-SA-access} before attaching SA credential & per-resource, as the service account \\
\bottomrule
\end{tabularx}
\end{table}

\subsection{The User\,$\rightarrow$\,Service-Account Check}

The most security-sensitive authorization decision is the User$\rightarrow$SA case
(Flow~3). Here a human triggers a tool that executes under a service account. If the
gateway forwarded the SA credential without checking, any user who could reach the tool
would inherit the service account's full privileges---a confused-deputy
escalation~\cite{hardy1988}. The gateway therefore performs an explicit entitlement check
(``is this user authorized to act through this service account?'') \emph{after} resolving
the user from their session and \emph{before} attaching the credential. The decision and
both identities are written to the audit log, so every SA-backed action is traceable to the
human who initiated it.

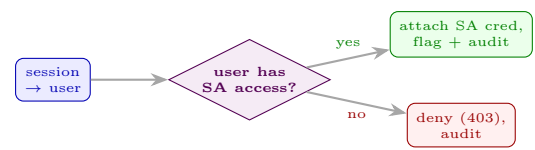
\begin{figure}[h]
\centering
\begin{tikzpicture}[x=1cm,y=1cm,
  d/.style={diamond,aspect=2,draw=cpurp,fill=clpurp,text=cpurp,
    font=\tiny\bfseries,align=center,inner sep=1pt},
  b/.style={rectangle,rounded corners=3pt,minimum height=0.55cm,
    font=\tiny,align=center,draw,fill=white},
  ok/.style={b,draw=cgreen,fill=clgreen,text=cgreen},
  no/.style={b,draw=cping,fill=clping,text=cping},
  arr/.style={-Stealth,thick,gray!70},
]
\node[b,draw=cblue,fill=clblue,text=cblue] (s) at (0,0) {session\\$\rightarrow$ user};
\node[d] (q) at (2.6,0) {user has\\SA access?};
\node[ok] (y) at (5.4,0.6) {attach SA cred,\\flag + audit};
\node[no] (n) at (5.4,-0.6){deny (403),\\audit};
\draw[arr] (s)--(q);
\draw[arr] (q)-- node[font=\tiny,text=cgreen,above]{yes} (y);
\draw[arr] (q)-- node[font=\tiny,text=cping,below]{no} (n);
\end{tikzpicture}
\caption{The User$\rightarrow$SA authorization gate. The entitlement check runs after the
session resolves to a real user and before any service-account credential is attached.}
\label{fig:authzgate}
\end{figure}

\section{Clients, Connectors, and Deployment}
\label{sec:deploy}

\subsection{Four Categories of Client}

The gateway is consumed by four client categories, each spanning different personas:

\begin{itemize}
  \item \textbf{Web clients} (browser-based AI chat UIs) --- user persona, via
    Authorization~Code~+~PKCE.
  \item \textbf{Desktop clients} (native AI apps) --- user persona, also via
    Authorization~Code~+~PKCE.
  \item \textbf{Custom-SDK clients} (built on agent frameworks such as CrewAI, the Claude
    Code SDK, the OpenAI Agents SDK, or LangChain, using Streamable-HTTP MCP transport)
    --- \emph{both} user and non-user personas, since these frameworks drive interactive
    sessions and unattended batch jobs alike.
  \item \textbf{Low-code / no-code clients} (workflow builders) --- \emph{both} personas,
    for the same reason.
\end{itemize}

\begin{table}[h]
\caption{Client categories and the personas they exercise.}
\label{tab:clients}
\small
\begin{tabularx}{\columnwidth}{Xcc}
\toprule
\textbf{Client category} & \textbf{User} & \textbf{Non-user} \\
\midrule
Web (chat UIs)          & \checkmark & \textemdash \\
Desktop (native apps)   & \checkmark & \textemdash \\
Custom SDK (agent fwks) & \checkmark & \checkmark \\
Low/no-code (workflows) & \checkmark & \checkmark \\
\bottomrule
\end{tabularx}
\end{table}

\subsection{Enterprise-Wide Connectors}

A practical contribution of the deployment is \emph{shared enterprise connectors}: a
single web connector and a single desktop connector for the gateway, configured once and
distributed to the entire workforce. An employee installs the connector and immediately
has governed access to every MCP server the gateway fronts, authenticated through corporate
SSO, with no per-server setup. This is what turns a back-end architecture into something
an entire organization actually uses---and it is only possible because authentication is
centralized at the gateway.

\subsection{Perimeter Evolution: From Edge to Tunnel}

Exposing the gateway to web clients raised a network-security question with two answers,
and we evolved from one to the other.

\textbf{Starting point --- CDN / WAF / edge route.} We initially fronted the gateway with
a CDN, a Web Application Firewall (with rules to block SQL-injection-style payloads), and
an edge route with a per-vendor IP allowlist. This works but is operationally heavy: every
new client vendor requires an allowlist change, and the WAF must be tuned to avoid blocking
legitimate MCP traffic.

\begin{figure}[h]
\centering
\begin{tikzpicture}[x=1cm,y=1cm,
  n/.style={rectangle,rounded corners=3pt,minimum width=1.5cm,minimum height=0.6cm,
    font=\tiny\bfseries,align=center,draw,fill=white},
  arr/.style={-Stealth,thick},
]
\node[font=\tiny\bfseries,text=cping] at (1.6,2.4){Before: edge perimeter};
\node[n,draw=cblue,fill=clblue,text=cblue]  (e1) at (0,1.6){Client};
\node[n,draw=cping,fill=clping,text=cping]  (e2) at (1.7,1.6){CDN+WAF};
\node[n,draw=cping,fill=clping,text=cping]  (e3) at (3.4,1.6){IP allow};
\node[n,draw=cgreen,fill=clgreen,text=cgreen](e4) at (5.1,1.6){Gateway};
\draw[arr,gray](e1)--(e2);\draw[arr,gray](e2)--(e3);\draw[arr,gray](e3)--(e4);

\node[font=\tiny\bfseries,text=cgreen] at (1.6,0.4){After: private tunnel};
\node[n,draw=cblue,fill=clblue,text=cblue]  (t1) at (0,-0.4){Client};
\node[n,draw=cgreen,fill=clgreen,text=cgreen](t2) at (2.6,-0.4){Tunnel};
\node[n,draw=cgreen,fill=clgreen,text=cgreen](t3) at (5.1,-0.4){Gateway};
\draw[arr,cgreen](t1)--(t2);\draw[arr,cgreen](t2)--(t3);
\node[font=\tiny,text=gray] at (2.6,-1.1){no public ingress, no IP allowlist};
\end{tikzpicture}
\caption{Perimeter evolution. The tunnel removes per-vendor IP allowlisting and public
ingress while leaving per-request token validation unchanged.}
\label{fig:perimeter}
\end{figure}
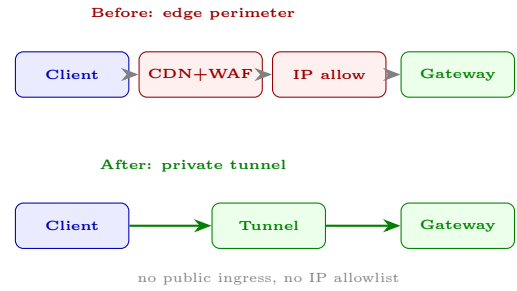

\textbf{Direction --- private MCP tunnels.} We are migrating to private MCP
tunnels~\cite{mcptunnel}, in which the client establishes an outbound tunnel to the gateway
rather than the gateway exposing a public ingress. This removes the need for per-vendor IP
allowlisting and shrinks the public attack surface. Critically, the tunnel does not change
the auth model: every request inside the tunnel still carries a fresh token that the
gateway validates independently. The tunnel secures \emph{transport}; the token secures
\emph{authorization}. Figure~\ref{fig:perimeter} contrasts the two.

\section{Discussion and Lessons Learned}

\textbf{The hard part was organizational, not cryptographic.} The OAuth standards
involved---PKCE, device code, ROPC, RFC~8693---are mature and well-implemented by
enterprise SSO products. The genuinely hard work was getting every server team to route
through the gateway instead of rolling their own auth, enabling token exchange on the SSO
(a feature the identity team had never previously needed), and agreeing on the
non-user-cannot-impersonate-user boundary.

\textbf{The two-axis model is the reusable idea.} Teams initially saw a bewildering list
of auth mechanisms. Reframing it as persona~$\times$~credential turned an unbounded problem
into a small matrix that a new server team can place themselves into in minutes.

\textbf{Centralization pays for itself at offboarding and audit.} The first time a security
review asked ``what did every agent do on behalf of every user last month,'' the unified
gateway log answered in one query---something that was impossible in the per-server world.

\textbf{Tool search matters as much as auth.} Aggregating dozens of servers behind one
endpoint would flood an agent's context with hundreds of tool schemas. Gateway-side tool
search keeps the agent's working set small, which is a precondition for the single-endpoint
model to be usable at all.

\section{Related Work}

\textbf{MCP security.} Recent work surveys enterprise MCP risks and mitigation
frameworks~\cite{narajala2025}, catalogues content-injection and overstepping-agent
adversaries~\cite{errico2025}, and measures ecosystem registry-vetting
weaknesses~\cite{ligao2026}. A concurrent line addresses server-side zero-trust
authorization and permission-filtered tool discovery~\cite{zerotrustmcp}. These works
identify the threat space; our contribution is a deployed gateway architecture that unifies
the heterogeneous \emph{authentication} layer those works assume exists.

\textbf{Zero-trust and identity.} NIST SP~800-207~\cite{nist800207} formalizes zero-trust;
BeyondCorp~\cite{beyondcorp} demonstrates enterprise-scale deployment. We apply these
principles at the MCP boundary, validating every request regardless of tunnel
establishment.

\textbf{OAuth standards.} Our design composes token introspection~\cite{rfc7662}, token
exchange~\cite{rfc8693}, PKCE~\cite{rfc7636}, and resource indicators~\cite{rfc8707} into a
single coherent gateway flow. Agent-identity delegation via the \texttt{act} claim follows
emerging identity-for-AI guidance.

\textbf{API gateways.} Service meshes and API gateways~\cite{servicemesh} terminate auth at
the HTTP boundary but do not address MCP-specific concerns: the two-persona-single-endpoint
requirement, downstream credential heterogeneity, or tool-search aggregation.

\section{Conclusion}

Enterprise MCP adoption outran governance: dozens of independently built servers, each with
its own auth, produced a landscape no one could authorize, audit, or offboard cleanly. We
described a production gateway that resolves this with a single authentication and
aggregation layer. The reusable ideas are a two-axis authentication model
(persona~$\times$~credential, one endpoint), a gateway layer offering three SSO grants and
three token-provisioning models (BYOT, GYOT, full OAuth via RFC~8693), and three
generalized identity flows---including the easily-mishandled User$\rightarrow$Service-Account
case. The architecture is in production across web, desktop, custom-SDK, and low-code
clients, fronting dozens of MCP servers through enterprise-wide shared connectors, and is
migrating from an edge perimeter to private tunnels without changing its per-request token
model. We believe the two-axis model and the gateway-issues/server-verifies split are
directly transferable to any organization facing the same MCP governance crisis.

\bibliographystyle{plain}

\end{document}